\documentclass[%
 aip,
 amsmath,amssymb,
 reprint,%
]{revtex4-1}

\usepackage{graphicx}% Include figure files
\usepackage{dcolumn}% Align table columns on decimal point
\usepackage{bm}% bold math
\usepackage{amsmath}
\usepackage{color}
\usepackage{epstopdf}
\usepackage{float}
\usepackage[utf8]{inputenc}
\usepackage[T1]{fontenc}
\usepackage{mathptmx}

\begin{document}

%\preprint{AIP/123-QED}

\title{Investigating microwave loss of SiGe using superconducting transmon qubits}
%\thanks{Footnote to title of article.}

\author{Martin. Sandberg}
\email{martinos@us.ibm.com}
\author{Vivekananda P. Adiga}
\author{Markus Brink}
\author{ Cihan Kurter}
\author{ Conal Murray}
\author{ Marinus Hopstaken}
\author{John Bruley}
\author{Jason Orcutt}
\author{ Hanhee Paik }
\affiliation{IBM Quantum, IBM T. J. Watson Research Center, Yorktown Heights, NY USA 10598}%Lines break automatically or can be forced with \\

\newcommand{\ket}[1]{\left|#1\right>}
\newcommand{\bra}[1]{\left<#1\right|}
\newcommand{\bracket}[2]{\left<#1|#2\right>}
\newcommand{\Bracket}[3]{\left<#1|#2|#3\right>}
\newcommand{\expvalue}[2]{\left<#1|#2|#1\right>}

\begin{abstract}

Silicon-Germanium (SiGe) is a material that \textcolor{black}{possesses} a multitude \textcolor{black}{of} applications ranging from transistors to eletro-optical modulators and quantum dots. The diverse properties of SiGe \textcolor{black}{also} make it attractive  \textcolor{black}{to implementations involving} superconducting quantum computing. Here we \textcolor{black}{demonstrate the fabrication of transmon quantum bits on SiGe layers and investigate the} microwave loss properties of SiGe at cryogenic temperatures and single photon microwave powers. We find \textcolor{black}{relaxation} times \textcolor{black}{of} up to 100 $\mu$s, corresponding to a quality factor Q above 4 M for large pad transmons. The high Q values obtained indicate that the SiGe/Si heterostructure is compatible with state of the art performance of superconducting quantum circuits.   
\end{abstract}

\maketitle

\begin{figure}[tb]
	\includegraphics[width=8.5cm]{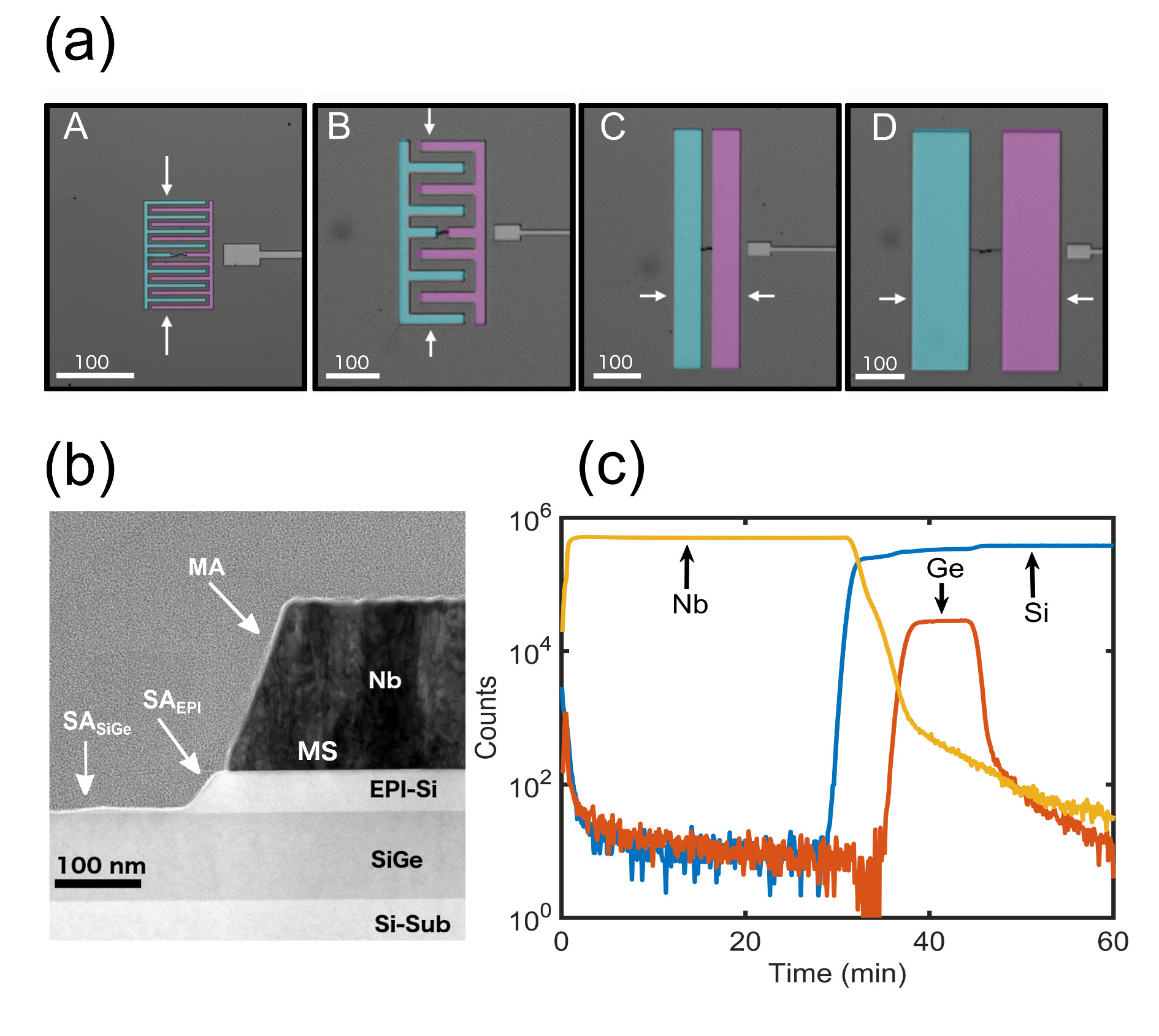}
	\caption{\label{fig_1}(a) \textcolor{black}{Top-down micrographs} of the transmon qubit devices used \textcolor{black}{in} the experiments. Four different device designs\textcolor{black}{, labeled} Model A, B, C and D\textcolor{black}{, are} shown in the sub-panels. Model A and B have interdigitated capacitor pads with gaps of 5 and 20 $\mu$m respectively. Model C has coplanar plates with \textcolor{black}{a} 20 $\mu$m \textcolor{black}{wide} gap and model D has coplanar plates with a gap of 70 $\mu$m.  In the figures the two capacitor pads of the transmon qubits have been false colored. \color{black} The arrows indicate the line of cross section used when modeling the surface participation. \color{black} (b) \textcolor{black}{Cross-sectional transmission electron microscope (TEM) image} of the device stack. \textcolor{black}{Key interfaces within the qubit geometry, such as the metal-air (MA), metal-substrate (MS), and substrate-air (SA) interfaces to the epi-Si and SiGe layers, are highlighted.} The Nb film \textcolor{black}{thickness is approximately} 190 nm, the epitaxial Si capping layer $\approx$ 50nm, \textcolor{black}{and} SiGe layer $\approx$ 100 nm. (c) Secondary ion mass spectrometry (SIMS) measurements performed on the device stack. The SIMS measurements suggest that the composition \textcolor{black}{ratio} of \textcolor{black}{Si:Ge} is 82:18 where as TEM- Energy Dispersive X-Ray Analysis (TEM-EDX) \textcolor{black}{suggests a ratio} of 87:13, both close to the desired value of 85:15.    } 
\end{figure}

Over recent years there \textcolor{black}{have} been significant \textcolor{black}{developments within} superconducting quantum processors. Currently small processors containing a few tens of qubits have been demonstrated \cite{IBMQX}. The operational fidelity of the processors are ever improving\cite{Preskill2018,Kjaergaard2019,Wendin_2017,Gambetta_rev2017}. It is realistic to believe that quantum algorithms will soon be able to outperform their classical counterparts \cite{Bravyi308} in several areas. However, to achieve fault tolerant quantum computing a multitude of problems needs to be addressed. A major limitation of quantum processors based on superconducting circuits is \textcolor{black}{the} relatively short coherence time of the quantum bits (qubits). By improving the intrinsic coherence properties of the qubits the overhead for operating a fault tolerant logical qubit can be reduced\cite{Fowler2012}. It is therefore of utmost importance that the coherence properties of the physical qubit are not degraded when new components or materials are incorporated in the quantum processor. 

Superconducting quantum circuits have traditionally been fabricated on either undoped silicon substrates or on sapphire substrates in order to maintain high coherence. There is significant interest in expanding \textcolor{black}{the} functionality of superconducting quantum \textcolor{black}{circuits} by incorporating new materials and structures \cite{nakamura_hybrid_2020}\cite{Petersson_gatemon_2018} \cite{oliver_gatemon_2018}. One material of great interest for a number of quantum  applications is Ge or Ge intermixed with silicon (Silicon-Germanium SiGe). Ge and SiGe \textcolor{black}{have} been incorporated in applications ranging from Josephson field effect transistors \cite{Silvano_2019} to spin qubits \cite{Seigo_2018} and recently it has been proposed that SiGe can provide a pathway for on-chip optical to microwave transduction \cite{Orcutt_2020}.  Here we investigate the coherence properties of transmon qubits\cite{Koch2007} fabricated on a Si substrate where a heterostructure of SiGe capped by an additional layer of epitaxial silicon (epi-Si) has been added. The possibility to combine SiGe technology with highly coherent superconducting quantum circuits could have significant \textcolor{black} {implications} for the development of quantum devices and applications. 

The main question we are trying to answer is if high coherence superconducting quantum circuits can be fabricated in conjunction with the growth of the Si/SiGe/Si stack. In order to test this \textcolor{black}{hypothesis}, we fabricate transmon devices with four different capacitor pad designs, as shown in FIG \ref{fig_1}a, on top of the heterostructure (sample G) and compare them to devices fabricated on our standard bare Si substrate (sample S). The capacitors pads are engineered to have varying amounts of electric field stored in the different layers and interfaces \cite{Gambetta2017}. The transmons are fabricated in a planar circuit geometry where each qubit has an on-chip Coplanar Waveguide (CPW) resonator for state readout. Each of the four designs are fabricated on 4$\times$8mm$^2$ chips with four identical qubits per chip. The readouts are addressed through a common feed line using hanger style capacitive couplings. 

The starting substrate for \textcolor{black}{sample} fabrication is a 750 $\mu$m thick high resistivity ($>$1 kOhm-cm) 200 mm silicon wafer. The native oxide is removed just prior to SiGe growth through \textcolor{black}{an} in-situ 1050$^{\circ}$C wafer anneal. A 100 nm thick SiGe layer with a Ge mole fraction of approximately 15\% is epitaxially grown using Rapid Thermal Chemical Vapor Deposition (RTCVD) \textcolor{black}{at a temperature of $650^{\circ}$C. } \textcolor{black}{This} composition is \textcolor{black}{commensurate with application to optical to microwave transduction that minimizes refractive index contrast and sensitivity to loss associated with scattering and absorption} \cite{Orcutt_2020} The layer is limited to this thickness in order to minimize formation of defects in the SiGe \cite{Hartmann_2011}. \textcolor{black}{X-ray diffraction analysis of the SiGe film (not shown) confirmed a fully strained layer corresponding to a Ge fraction of approximately 16.5\%.} The SiGe is capped with a 50 nm layer of epitaxially grown Si. The effective dielectric constant for the Si$_{1-x}$Ge$_x$ layer is approximated through the expression\cite{Schaffle2001}  $\epsilon_{SiGe}=\epsilon_{Si} + 4.5x$  which for $x= 15$\% mole fraction gives $\epsilon_{SiGe} \approx$ 12.35. To form the superconducting circuitry a 200 nm thick layer of Niobium is sputtered on top of the Si/SiGe/epi-Si stack. The qubit capacitor pads and readout resonators are patterned using a subtractive \textcolor{black}{process that incorporates} optical lithography and Reactive Ion Etch (RIE). As can be seen in the \textcolor{black}{cross-sectional, transmission electron microscope image}, FIG. \ref{fig_1}(b),  the RIE process creates an over-etch of approximately 50 nm \textcolor{black}{which leads to a partially exposed SiGe surface} where the Nb has been removed.  As a final step an Al/AlO$_x$/Al Josephson junction is placed between the capacitor pads. The \textcolor{black}{Josephson} junction is fabricated using a Dolan bridge technique based on a PMMA/MMA resist stack and e-beam lithography. The aluminum is evaporated at two different angles with an oxidation step prior to the second deposition. Reference samples are formed on an identical Si substrate but without the presence of the SiGe and epi-Si layers.

A major source of loss \textcolor{black}{in} superconducting qubits is attributed to Two Level Systems (TLS) that couple to the electric field of the qubits \cite{Burnett2019,Schlor2019}. The physical origin of the TLS's is not fully understood but evidence suggests that they are  primarily located at device surfaces and interfaces \cite{gao2008}. More precise pinpointing of the TLS locations can be achieved using electrostatic gates\cite{Bilmes_2020}. Such measurements indicate that the dielectric surfaces close to conductor edges, as well as the Josephson junction are areas of high TLS density.  To estimate how much \textcolor{black}{such regions} of the circuit \textcolor{black}{may contribute} to the qubit loss, we determine the amount of electric field energy stored in \textcolor{black}{each} region. The electric field energy fraction is referred to as \textcolor{black}{the} filling factor or participation ratio.  Due to the high aspect ratio of the \textcolor{black}{interfacial} layers for a typical transmon \textcolor{black}{design} (a few nm thickness  and up to a few mm in the plane), and the fact that the electric field diverges at the edges of the conductors, calculations of \textcolor{black}{the} filling factor are not straightforward. Several methods have been suggested for extracting surface participation \cite{Gambetta2017,Wang2015,Wenner2011,sandberg2012,Woods2019,Murray2018}. Here we calculate filling \textcolor{black}{factors} $F_L$ for the different interfaces and the SiGe and epi-Si layers $L$ using a 2D electromagnetic Finite Element (FEM) solver to extract the static electric field distribution for the qubit designs shown in FIG \ref{fig_1}(a). For each design we create a cross\textcolor{black}{-}sectional model\textcolor{black}{;} for interdigitated capacitors (IDC) qubits (design A and B), a representative cross\textcolor{black}{-}section is taken across the fingers and for parallel \textcolor{black}{plates} (design C and D)  across the pads \color{black} as indicated by arrows in FIG. \ref{fig_1} (a) \color{black}. In the FEM model we assign a static differential voltage to the two capacitor pads of +1 V and -1 V and 0 V to any ground electrodes \color{black} (further details on assumed interface layer thicknesses and dielectric constants are provided in Table \ref{SiGe_fill}).\color{black}

\begin{equation}
F_L=\frac{\int_{L}\vec{E}\cdot \vec{D}}{\int_{V}\vec{E} \cdot \vec{D}}
\label{eq1}
\end{equation}
     
 Using the expression in \textcolor{black}{Eq.} (\ref{eq1}) we \textcolor{black}{can solve for} the filling \textcolor{black}{factors of} the epitaxial Si capping layer\textcolor{black}{,} the SiGe layer as well as the other surfaces and interfaces of the device
\textcolor{black}{.} The values are summarized in \textcolor{black}{Table} \ref{SiGe_fill}. The combined surface loss can then be expressed as:

 \begin{equation}
 \frac{1}{Q_{tot}} =  \sum_{L}F_{L}\tan(\delta_{L})+\delta_{bg}
 \label{eq2}
 \end{equation}
 
where \textcolor{black}{$Q_{tot}$ refers to the measured qubit quality factors,} $L\in \{ \mathrm{epi, SiGe, MS, MA, SA_{epi}, SA_{SiGe}, Si} \}$ , \textcolor{black}{$\tan(\delta_{L})$ the corresponding loss tangents} and \color{black} $\delta_{bg}$ \color{black}
\textcolor{black}{represents} all other sources of loss, such as radiation, quasiparticles\textcolor{black}{,} etc.

\begin{table}[tb]
	\begin{tabular}{|c|c|c|c|c|c|c|}
		\hline
		Design & Epi-Si & SiGe & MS & MA & SA$_{SiGe}$ & SA$_{EPI}$\\
		            & ($\times 10^{-2}$) &($\times 10^{-2}$) &($\times 10^{-3}$)&($\times 10^{-4}$)& ($\times 10^{-4}$) &  ($\times 10^{-4}$)  \\
		\hline \hline
		A           & 2.4           & 6.68                              & 2.40                      & 4.08                  & 2.37                     & 1.57 \\
		B           &0.74          & 2.20                              & 0.76                    &1.02                    &0.78                   & 0.39 \\
		C           & 0.41         & 1.24                              & 0.42                    & 0.54                 & 0.44                 & 0.21\\
		D           & 0.18         & 0.56                              & 0.19                    &  0.21                &0.20                   &0.08  \\
		\hline
	\end{tabular}
	\caption{\label{SiGe_fill} Filling \textcolor{black}{factors of} the SiGe layer, the epitaxial \textcolor{black}{Si} capping layer, the Metal-Substrate (MS), Metal-Air (MA) and the two Substrate-Air (SA) interfaces SA$_{SiGe}$ and SA$_{EPI}$. The filling factors are calculated using a FEM solver \textcolor{black}{employing} a 2D \textcolor{black}{cross-sectional geometry} of the samples. The relative dielectric constant of Nb$_2$O$_5$ can vary vastly, in the calculations we are assuming $\epsilon_r = 30$ \cite{PIGNOLET199518} and a thickness of 5nm for MA, $\epsilon_r = 3.9$ and a thickness of 2nm for MS and the two SA layers. }	
\end{table}
     
\begin{table*}[t]
	\begin{tabular}{|c|c|c|c|c|c|c|c|c|}
		\hline
		 & \multicolumn{2}{c|}{A} & \multicolumn{2}{c|}{B}  & \multicolumn{2}{c|}{C} &  \multicolumn{2}{c|}{D}  \\
		
		  & \multicolumn{2}{c|}{T$_1$ ($\mu$s)/f$_Q$ (GHz) }& \multicolumn{2}{c|}{T$_1$ ($\mu$s)/f$_Q$ (GHz) }  &  \multicolumn{2}{c|}{T$_1$ ($\mu$s)/f$_Q$ (GHz) } & \multicolumn{2}{c|}{T$_1$ ($\mu$s)/f$_Q$ (GHz) }  \\
		\hline 
		 Qubit  &    G     &    S              &      G       &       S       &   G             &    S             &    G                           &  S  \\
		Q$_1$ & 18/5.258 & 23/4.601    & --/5.441 & 48/4.638  &80/5.567   &   90/4.622   &112/5.472                   & 73/4.774    \\
		Q$_2$ & 26/5.428 & 18/4.645   & 41/5.521&   55/4.786 &71/5.475    &   79/4.657   &20$^{\dagger}$/5.195 & 95 /4.722  \\
		Q$_3$ & 28/5.731 & 13/4.670    & 58/5.687&  53/4.762 &95/5.346   &  89/4.674    &107/5.489                   &98 /4.786    \\
		Q$_4$ & 27/5.552 & 22/4.489   & 57/5.557&  40/4.906 &130/5.460 &  71/4.735     &114/5.635                   & 81/4.686    \\
		\hline
		mean   & 24.7/5.492 & 19/4.602 &52/5.552 & 49/4.773 & 94/5.462  &  83/4.672    &110/5.448                    & 87/ 4.742\\
		\hline
	\end{tabular}
	\caption{\label{t1_fq}  Measured energy relaxation times and frequencies for each qubit of sample G (SiGe) and sample S (Si). The T$_1$ times are measured over a period of 3 hours with approximately 20 measurements each. \color{black}The standard deviation of the T$_1$ measurements in a time series is typically in the range \textcolor{black}{of} 5-30\% \color{black}.  The For Q$_1$ on chip B the qubit pulses were not calibrated correctly so no data was acquired. $^{\dagger}$This outlier value is somewhat surprising and we have excluded it from the calculation of the mean value. It is possible that  Q$_2$ on chip D was temporarily strongly coupled to a TLS that reduced the T$_1$ time significantly. }	
\end{table*}

\begin{figure}[tb]
	\includegraphics[width=8.5cm]{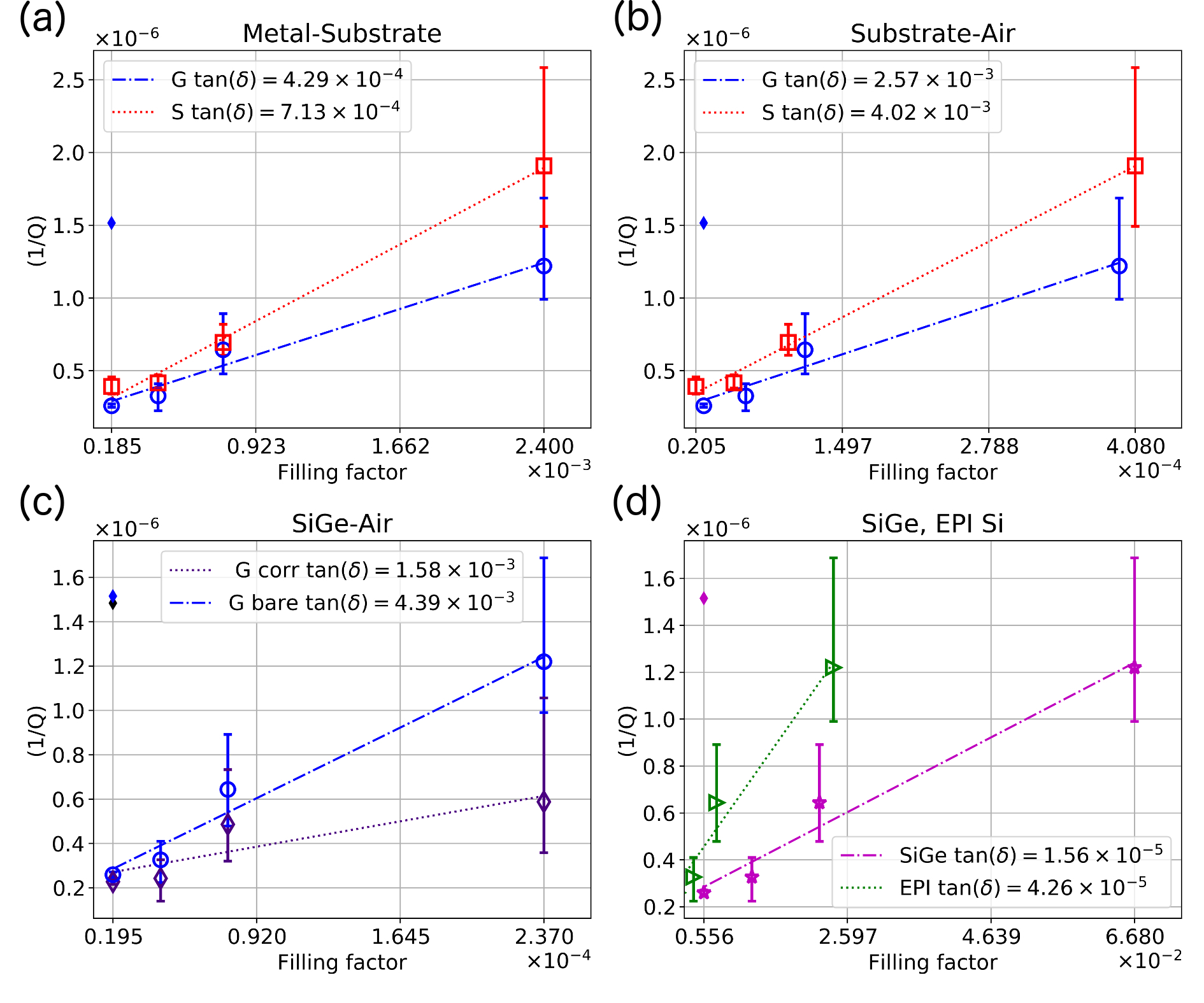}
	\caption{\label{fig_2}  Extracted bounds of the loss tangent for six layers and interfaces. The loss \textcolor{black}{corresponding to the reciprocals of the measured quality factors (1/Q)} for the different qubit designs \textcolor{black}{is}  plotted as a function of participation for the different layers SiGe, epi-Si, MS, SA$_{SiGe}$ and SA$_{epi-Si}$. The slope of the linear fit gives an upper bound on the loss tangent for the different interfaces. \bf(a) \rm Metal-substrate interface for sample G and S, \bf (b) \rm Substrate-Air interface for sample S and G. \bf(c) \rm  For sample G\textcolor{black}{,} the \textcolor{black}{sidewall components of the} SA interface are all Si and the trench \textcolor{black}{bottom} is SiGe.  By subtracting the sidewall loss we can get a bound for the SiGe-Air interface. \bf(d) \rm  Bounds on the loss tangents for the SiGe layer and the epi-Si layer respectively.  }
\end{figure}

The devices were packaged and then measured in a $^3$He/$^4$He dilution refrigerator with a base temperature of approximately 15mK. In our setup the input lines are heavily attenuated (in total 70 dB of attenuation distributed over the different thermal stages such that the added thermal noise only corresponds to a few mK). The outputs from the four chips are routed through an electro-mechanical microwave switch and amplified by a HEMT amplifier. The samples are protected from thermal noise coming from the HEMT input by two 4-12 GHz cryogenic isolators, placed at the base temperature of the cryostat. One chip (4 qubits) of each design was measured in a single cooldown. The frequencies and energy relaxation times (T$_1$) are shown in \textcolor{black}{Table} \ref{t1_fq} where we see the expected trend that qubit \textcolor{black}{designs} with closely spaced capacitor pads \textcolor{black}{exhibit} lower T$_1$. 
 
From the T$_1$ and qubit frequency (f$_Q$) we extract the quality factors \color{black}  Q$_{tot}=2\pi$f$_Q$ T$_1$. \color{black} The qubit frequencies are approximately 0.65 to 1.1 GHz higher for SiGe devices, however the small coupling to the readout resonators (< 50 MHz), large detuning (> 1.2 GHz) and a small resonator kappa (< 1 MHz) leads to a Purcell induced relaxation time of at least 550 microseconds for any of the qubits.  We therefore concluded that Purcell loss is not impacting the observed energy relaxation times in a significant manner. \color{black}  The total loss ($1/Q_{tot}$) for each design is plotted as a function of filling factor for the different interfaces in FIG. \ref{fig_2}. From a linear fit we extract an upper bound for the loss tangent of each interface or layer. The fits for the SiGe \textcolor{black}{and epi-Si layers correspond to upper bounds} on the loss tangent of  \textcolor{black}{$\tan(\delta_{SiGe})<1.56\times10^{-5}$ and $\tan(\delta_{epi})<4.26\times10^{-5}$} respectively (see FIG\textcolor{black}{.} \ref{fig_2}(d)). Even though these \textcolor{black}{values} are orders of magnitude \textcolor{black}{larger than that of} bulk Si ($< 3\times10^{-7}$) they \textcolor{black}{confirm} that high qubit quality factors \textcolor{black}{can be achieved in the presence of such layers}. It is also reasonable to assume that the actual loss of the SiGe and epi-Si could be significantly lower than these bounds, considering that devices formed on bare Si substrates typically do not outperform the devices measured here. As discussed previously the Nb overetch stopped just at the interface between the SiGe and the epi-Si. We hence divide the substrate-air (SA) interface into two parts\textcolor{black}{:} the Si sidewalls and the SiGe trench surface. If we use the extracted loss tangent for SA of sample S we can subtract the loss contribution of the sidewalls and get a more accurate bound on the SiGe-air interface loss. This is indicated in FIG. \ref{fig_2}(c). We find a bound on the loss tangent of 1.58$\times10^{-3}$ which \textcolor{black}{suggests} that the SiGe-Air interface is as good if not better than the Si-Air interface, \textcolor{black}{which could explain} why sample G \textcolor{black}{outperforms} sample S in a majority of \textcolor{black}{the} cases. 

\begin{figure}[tb]
	\includegraphics[width=8cm]{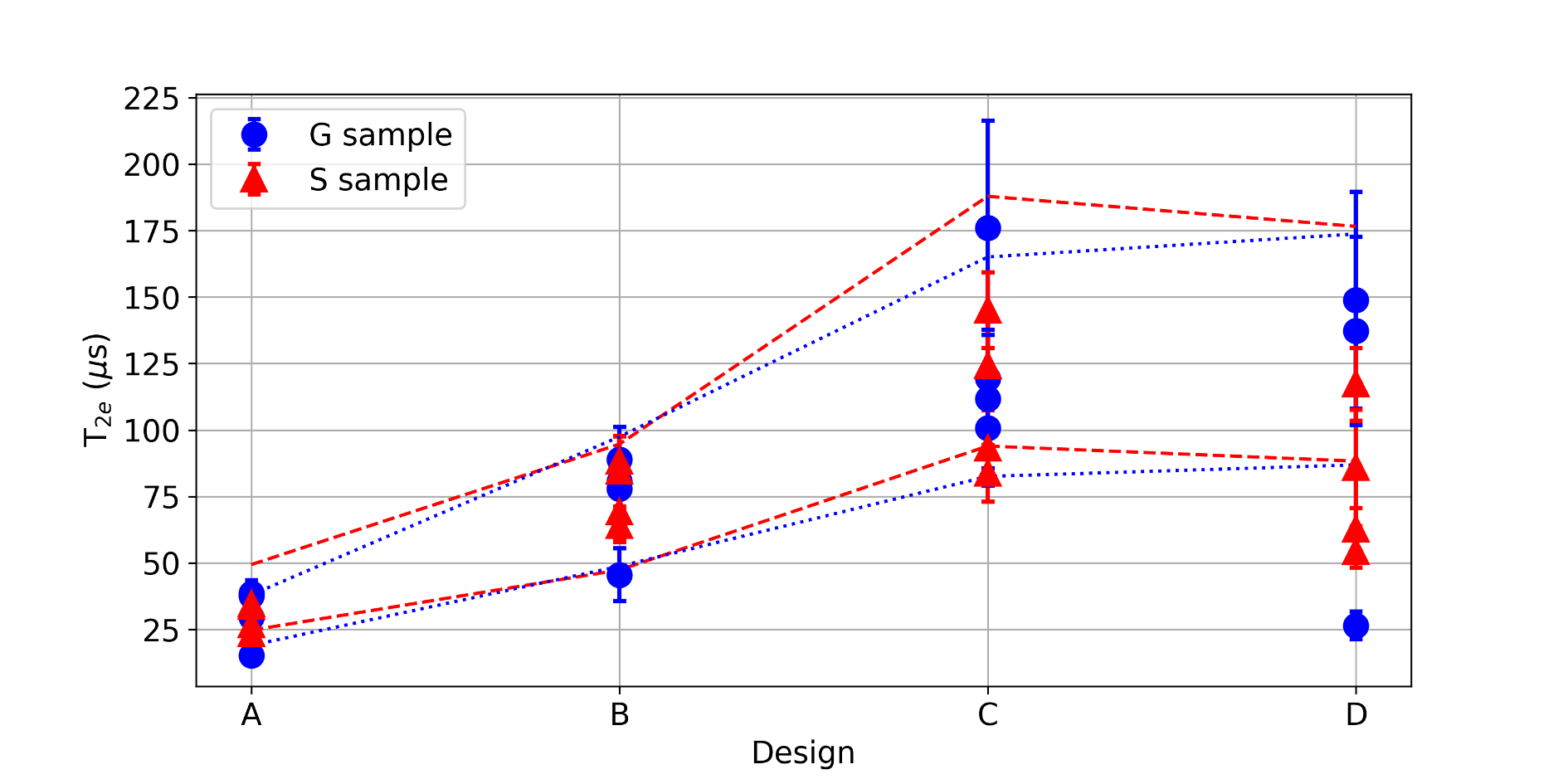}
	\caption{\label{fig_3}  Measured T$_{2e}$ \textcolor{black}{values} for all \textcolor{black}{of} the qubits. \color{black}The lines represent the mean T$_1$ and 2T$_1$ boundaries for the S \textcolor{black}{(dashed)} and G \textcolor{black}{(dotted) samples}. As is commonly observed for transmon qubits the T$_{2e}$ values \textcolor{black}{do} not reach the theoretical limit of 2T$_1$ but tend to \textcolor{black}{lie} between T$_1$ and 2T$_1$.  The mean T$_{2e}$/T$_1$ ratio for the SiGe sample is found to be \textcolor{black}{comparable to} the reference Si sample, suggesting that the \textcolor{black}{presence} of SiGe does not add any additional channels of depahsing over \textcolor{black}{those associated with the} bare Si substrate. \color{black}   }
\end{figure}

We also measured the T$_2$ echo times (T$_{2e}$) of the devices, \textcolor{black}{plotted in} FIG. \ref{fig_3}. We find that the T$_{2e}$ \textcolor{black}{values} tend to not be fully 2$\times$T$_1$ limited but are typically closer to T$_1$ indicating that our experimental setup also produces a significant fraction of external dephasing. To \textcolor{black}{determine} if the SiGe \textcolor{black}{layer} introduces any extra \textcolor{black}{sources} of depahsing we extract a T$_{2e}$/T$_1$ ratio averaged over all qubits for each substrate. For the devices fabricated on a pure Si substrate we \textcolor{black}{arrive at} a value of 1.365 \textcolor{black}{whereas} for the SiGe substrate the value is 1.37\textcolor{black}{, suggesting} that the SiGe \textcolor{black}{does} not contribute excess depahsing over \textcolor{black}{that from} pure Si substrates.      

To conclude, we have \textcolor{black}{ fabricated} high coherence transmon qubits on a stack of Si/Si$_{0.85}$Ge$_{0.15}$/epi-Si \textcolor{black}{layers, and} observe T$_1$ and T$_{2e}$  times in excess of 100 microseconds, on par with state of the art transmon devices. We find that even though these layers are relatively thick, \textcolor{black}{their} electric energy participation still scales in a fashion similar to surface participation for the different capacitor designs. By naively assigning all loss to one of the layers we can estimate an upper bound on the loss tangent. It is very likely that the extracted loss tangent is significantly lower, and that the dominant portion of the observed loss \textcolor{black}{arises from the typical} MS, SA and MA interface regions. The fact that high coherence devices can be \textcolor{black}{manufactured} on a Si/SiGe/Si material stack opens up \textcolor{black}{many possible applications where the incorporation} of SiGe can be integrated with superconducting quantum circuits.

\section{Data Availability Statement}
The data that support the findings of this study are available from the corresponding author upon reasonable request.

\begin{acknowledgements}
This work was funded by LPS/ARO under CQTS program, Contract Number W911NF-18-1-0022. The authors gratefully acknowledge \textcolor{black}{Dr. Madhana Sunder for X-ray diffraction characterization and} support in device fabrication by the IBM Microelectronics Research Laboratory and IBM Central Scientific Services.   
\end{acknowledgements}

\end{document}